\documentclass[aps,prl,twocolumn,showpacs,groupedaddress]{revtex4}
\usepackage[dvips]{graphicx}
\usepackage{hyperref}
\usepackage{amsmath}

\begin{document}

\title{Single InAsP/InP quantum dots as telecommunications-band photon sources}

\author{D.\ Elvira}
\author{R.\ Hostein}
\author{B.\ Fain}
\author{L.\ Monniello}
\author{A.\ Michon}
\author{G.\ Beaudoin}
\author{R.\ Braive}
\author{I.\ Robert-Philip}
\author{I.\ Abram}
\author{I.\ Sagnes}
\author{A.\ Beveratos}
\affiliation{CNRS - Laboratoire de Photonique et Nanostructures,
Route de Nozay, F-91460 Marcoussis, FRANCE}
\date{\today}

\begin{abstract}

The optical properties of single InAsP/InP quantum dots are investigated by spectrally-resolved and time-resolved photoluminescence measurements as a function of excitation power. 
In the short-wavelength region (below 1.45 $\mu$m), the spectra display sharp distinct peaks resulting from the discrete electron-hole states in the dots, while in the long-wavelength range (above 1.45 $\mu$m),
these sharp peaks lie on a broad spectral background. 
In both regions, cascade emission observed by time-resolved photoluminescence confirms that the quantum dots possess discrete exciton and multi-exciton states.
Single photon emission is reported for the dots emitting at 1.3 $\mu$m through anti-bunching measurements. 

\end{abstract}
\pacs{
     78.67.Hc Optical properties of Quantum dots ; 78.47.jd   Time resolved luminescence; 42.50.Ar  Photon statistics and coherence theory 
     } % end of PACS codes
 \maketitle

%%%%%%%%%%%%%%%%%%%%%%%%%%%%%%%%%%%%%%%%%%%%%%%%%%%%%%%%%%%%%%%%%%%
\section{Introduction}

The three-dimensional confinement of electrons and holes in semiconductor quantum dots gives rise to discrete electron-hole states and sharp absorption and emission lines, analogous to those in atomic systems \cite{Kastner1993}. 
These features have been exploited to produce quantum states of light, such as single photons \cite{Santori2001, Robert2001}, indistinguishable photons \cite{Santori2002a, Varoutsis2005, Laurent2005} and entangled photon pairs
\cite{Stevenson2006, Akopian2006,Dousse2010} that may be used in quantum communication protocols, such as quantum key distribution or quantum relays based on quantum teleportation \cite{Ekert1991, Bennett2000,Jennewein2000,Collins2005}. 
At the same time, quantum dots have been used as gain media in photonic crystal nanolasers \cite{Strauf2006}.
However, for highly-excited quantum dots placed inside photonic crystal nanocavities, it was found that the simple ``artificial atom'' model of the quantum dot, which successfully described the emission of one or two photons by the quantum dot in free space, could not adequately explain the emission of light by the dot into an apparently non-resonant nanocavity \cite{Henessy2007}. This cavity feeding required explicit consideration of multiply-excited states emitting into a broad quasi-continuum \cite{Winger09}.

To date most such photon sources and nanolasers have been fabricated with quantum
dots embedded in a GaAs matrix and thus emitting around 920 nm, while prospective
applications require sources operating in the telecommunications wavelength range,
particularly in the O- and C-bands, around 1.3 $\mu$m and 1.5 $\mu$m respectively. 
InAs/InP quantum dots can emit in these wavelength bands and are well suited as active
media in semiconductor optical amplifiers or ridge laser systems useful for telecommunications applications
\cite{Reithmaier2005}. 
However, attempts to grow such dots by Molecular Beam Epitaxy (MBE) did not give the
desired results, as growth on (001)-InP substrate generally leads to the formation
of quantum dashes or quantum wires \cite{Brault01, Garcia01}, while growth on a (311)-InP oriented substrate \cite{Paranthoen01} is not compatible with the standard processes used in the fabrication of photonic
devices such as microcavities.  
Use of Metal-Organic Chemical Vapor Deposition (MOCVD), on the other hand, has made
it possible to grow small InAsP/InP islands on a (100)-InP oriented substrate, as it
allows for the spontaneous formation of a two-dimensional wetting layer on which
small islands can grow \cite{Marchand06, Carlsson98, SaintGirons06}, while their
spectral distribution \cite{Michon08} or density \cite{Veldhoven09} can be adjusted
during growth. Transmission Electron Microscopy (TEM) studies on MOCVD-grown samples have shown that the islands are truncated pyramids of diamond-shaped cross-section with diagonals of the order of 30 nm to 40 nm at the top \cite{Michon08}.

Although these islands are small enough to have discrete electronic states, sharp spectral lines (identified as exciton and biexciton lines \cite{Takemoto2005, Kuroda07, Sek09, Veldhoven09}) are observed only in the
short wavelength side of the luminescence spectrum, from 1.3 $\mu$m up to 1.45 $\mu$m. 
By contrast, the emission spectrum around 1.55 $\mu$m generally exhibits a large
number of low intensity peaks lying on a broad and intense background, a feature
that may be interpreted as corresponding to a continuum of electron-hole states in
the InAsP island. 
This spectral region has not been explored up to now, and the nature of the states that contibute to it, whether continuous or discrete, has not yet been clarified.

In this paper we investigate the light emission properties of InAsP/InP islands and
show that they are indeed quantum dots whose discrete exciton states can provide
individual photons, while their multi-exciton states give rise to cascaded emission
that can feed a nanocavity and provide gain for nanolasers. 
The paper is organized as follows: We first describe the sample growth and the experimental setup. Then, we identify discrete exciton lines of two distinct quantum dots, emitting in each of the two spectral regions and study their saturation characteristics. Then, through time-resolved measurements, we show that both types of quantum dots have discrete multi-exciton states. Finally, we demonstrate single photon emission from a single quantum dot.

\section{Sample growth and experimental setup}

Samples were grown in a vertical-reactor low-pressure MOCVD system using hydrogen as the carrier gas and standard precursors (arsine, phosphine, and trimethylindium) \cite{Michon05}. 
The epitaxied InAsP islands were formed on a thick ($\sim$200 nm) InP buffer layer deposited on a (001)-oriented semi-insulating InP:Fe substrate. 
The island growth was obtained at 510 $^\circ$C by depositing 6.3 monolayers (ML) of
InAsP at growth rate of 0.36 ML.s$^{-1}$ and under a phosphine/arsine flow ratio of
30. Finally, a 63 nm thick InP capping layer was grown over the islands at a rate of
0.2 ML.s$^{-1}$. Such a growth sequence leads to the formation of InAsP islands with
an average height of 3.8 nm and a density of 15x$10^9$ cm$^{-2}$, sitting on a 1.5 nm-thick wetting layer, measured by TEM
experiments \cite{Michon08}. 
The average composition of the islands was measured to be InAs$_{0.8}$P$_{0.2}$, while the residual doping of the InP buffer layer was measured to be around $10^{16} e^- cm^{-3}$, indicating that the islands may be charged, containing one or more electrons.

Small numbers of islands were isolated by etching mesas using the following
technique. Layers of SiN and Polymethyl Methacrylate (PMMA) were deposited on top of
the sample. 500 nm up to 2 $\mu$m diameter holes were formed in the PMMA layer by
electron beam lithography. After deposition of a 40 nm thick layer of nickel
followed by a lift-off, the SiN layer was etched by reactive ion etching (with gaz mixture of SF$_6$ and CHF$_3$), to form bilayer pillars of SiN and Ni on the semiconductor. These pillars acted as a mask in the subsequent
semiconductor etching in an inductively coupled plasma $-$ reactive ion etching
machine (with gaz mixture of HBr and O$_2$). This whole process led to the formation
of 1.3 $\mu$m tall mesas in the semiconductor, with diameters ranging from 500 nm to 2 $\mu$m.

Time-resolved microphotoluminescence experiments were performed on the samples, placed in a liquid He flow cryostat, under pulsed excitation with 5 ps-long pulses at 80 MHz delivered by a Ti:Sa laser emitting at 840 nm. The excitation pulses were focused on the samples by a microscope objective (numerical aperture of 0.4) to a spot of a diameter of 5 $\mu$m. An incident power of 100 nW thus corresponds to a pulse energy of 125 nJ.cm$^{-2}$.
The island luminescence was collected by the same microscope objective and separated from the pumping laser by means of a dichroic mirror and an antireflection coated Si filter. 
The spontaneous emission was spectrally dispersed by a 0.5 m spectrometer and detected either by a cooled InGaAs photodiode array (Roper Scientific) or time-resolved superconducting single photon counters (SCONTEL) with a time resolution of 50 ps, a measured quantum efficiency of 3\% at 1.55 $\mu$m and dark count rates lower than 30 counts per second.
Lifetime measurements were obtained by recording the histogram of the time interval between a photon detection and the subsequent laser pulse using a LeCroy 725Zi oscilloscope. 
The second order auto-correlation function was measured with a standard Hanbury-Brown and Twiss setup: The collected
luminescence signal was split by a 50/50 fibered coupler and sent onto two single-photon detectors placed on each output of the beamsplitter. 
The auto-correlation function was deduced from the histogram of the time intervals between two single photon detection events. 
%The emission linewidths were obtained under continuous wave excitation (with a
%frequency-doubled Nd:YAG laser emitting at 532 nm) by measuring the first order
%autocorrelation function $g^{(1)}(\tau)$ with a Michelson interferometer placed
%between the microscope objective and the spectrometer \cite{Kammerer2002}.

\begin{figure}[!ht]
    \centering
     \includegraphics[width=7.5cm]{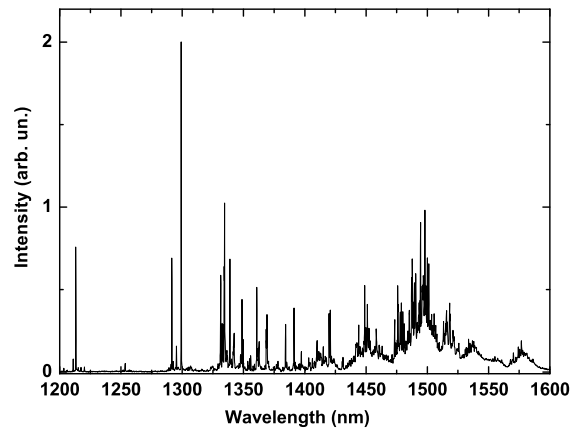}
%        \begin{tabular}{rr}
%        \includegraphics[width=4.5cm]{Fig1b.eps} &
%        \includegraphics[width=4.5cm]{Fig1c.eps} \\
%        \end{tabular}
    \caption{Typical low-temperature photoluminescence spectra observed on a 1 $\mu$m diameter mesa under pulsed excitation at 840 nm with an incident power of P$_{in}$ = 162 nW (cw equivalent power measured after the microscope objective, corresponding to 200 nJ/cm$^2$ per pulse).}
        \label{Fig:TypicalSpectrum}
\end{figure}

\section{Spectral characteristics}

The emission spectrum of the unprocessed sample (not shown here) consists of a broadband emission centered at 1.5 $\mu$m having full width at half-maximum of 0.13 $\mu$m and a long tail extending towards the shorter wavelengths, reflecting essentially the size statistics of the InAsP islands \cite{Hostein08}.
Figure \ref{Fig:TypicalSpectrum} presents a typical emission spectrum obtained under pulsed excitation at 4 K on a 1 $\mu$m diameter mesa containing approximately 120 islands, of which probably 75 \% are far enough from the mesa edge to luminesce properly.
The spectrum, which consists of the superposition of the spectra of several islands, displays distinct and sharp intense peaks in the short wavelength part (below 1.45 $\mu$m). As wavelength increases, the peaks become denser, reflecting the size statistics of the quantum dots. At the same time, the peaks get less intense while a broad background becomes increasingly strong as wavelength increases.  
These spectral features have already been reported by several other groups \cite{Takemoto2005, Kuroda07, Sek09, Veldhoven09} with similar, but not identical, systems, but their nature had not been investigated.

\begin{figure}[!ht]
    \centering
		\begin{tabular}{rr}
        \includegraphics[width=4.5cm]{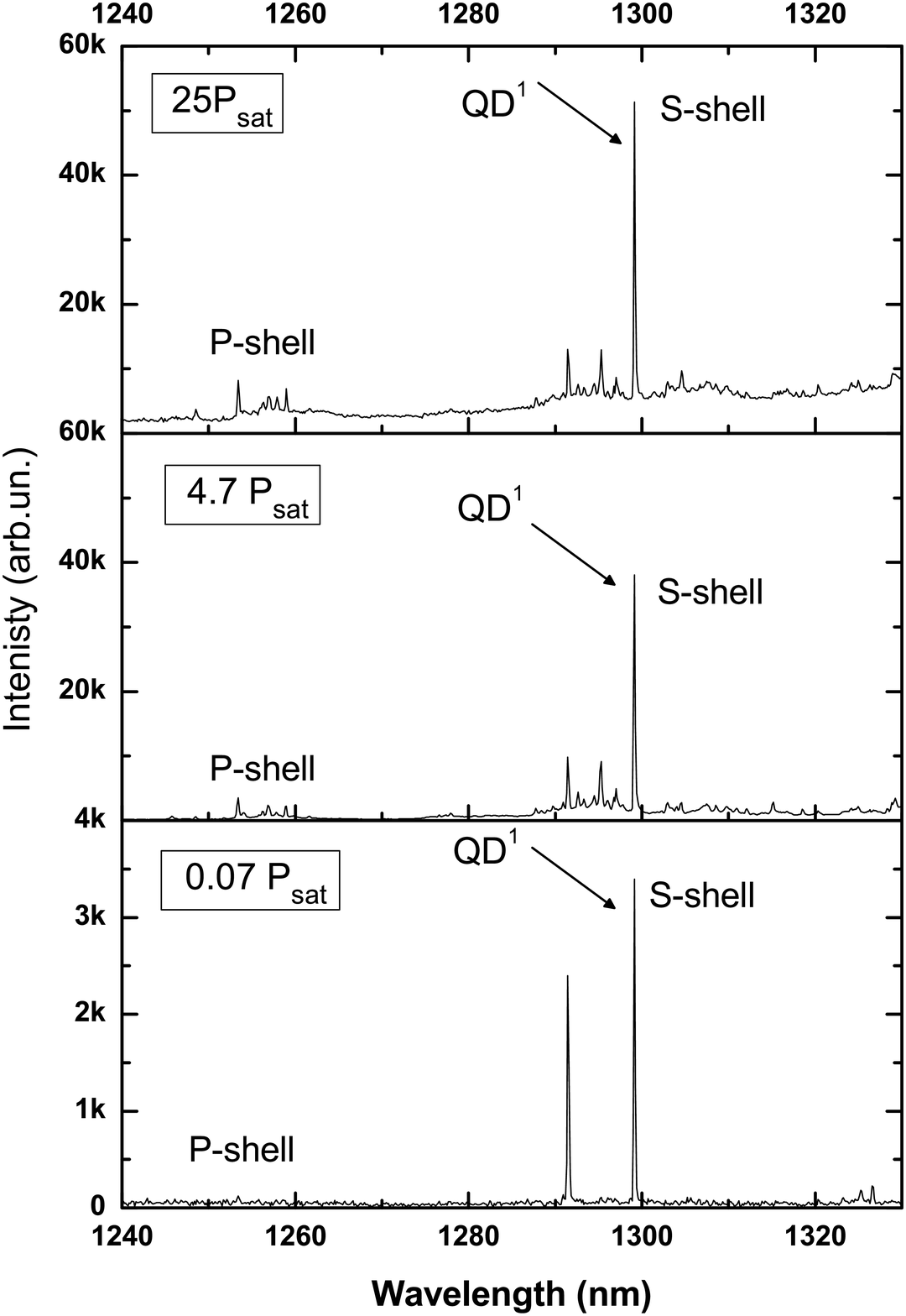} &
        \includegraphics[width=4.5cm]{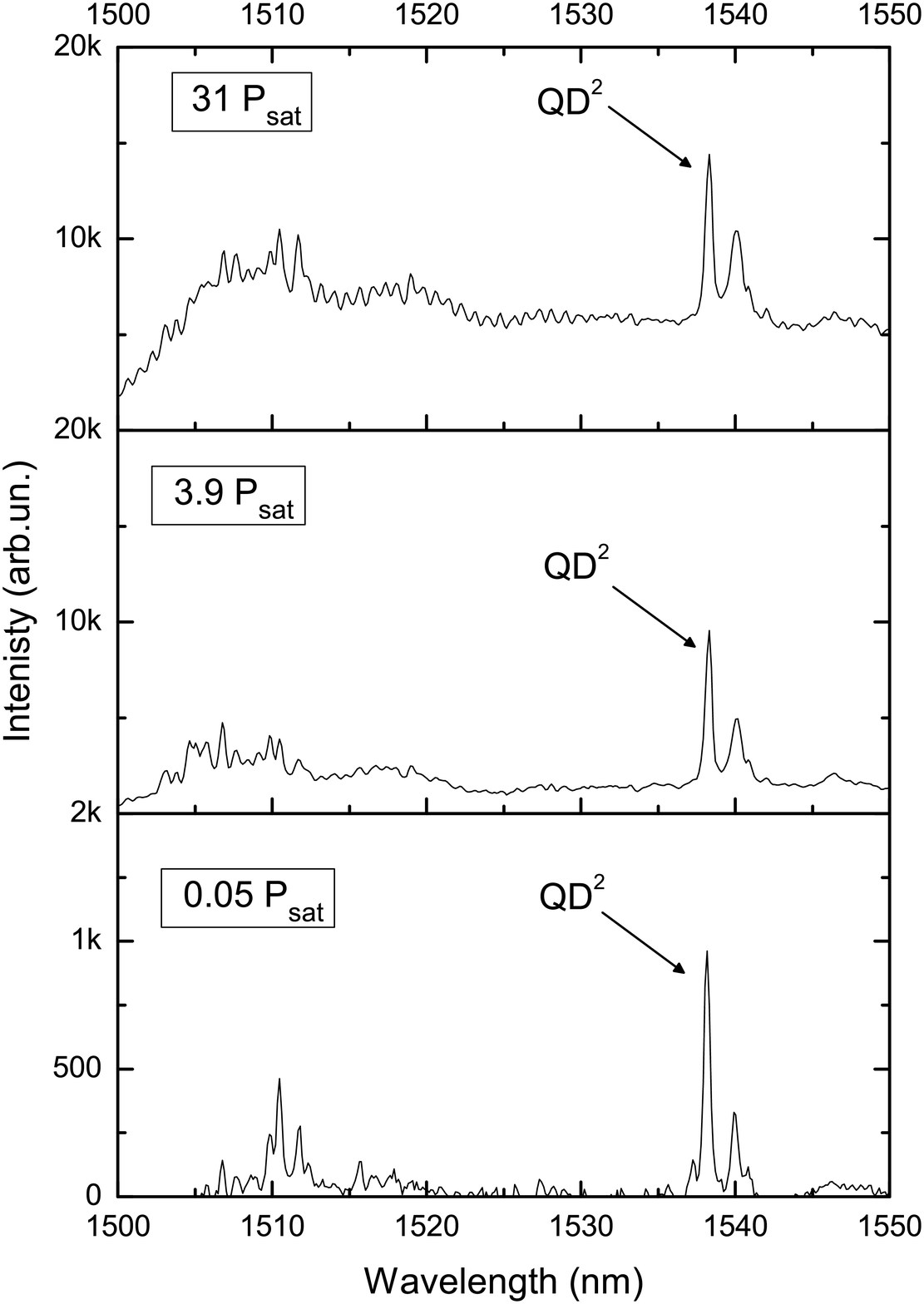} \\
        \end{tabular}
    \caption{Low-temperature spectra of the emission lines associated with two different InAsP/InP quantum dots labeled QD$^{1}$ (Left) and QD$^{2}$ (Right) obtained under pulsed excitation at 840 nm, at different incident powers. Incident powers are normalized to P$_{sat}=$ 620 nW for QD$^{1}$ and P$_{sat}=$ 180 nW for QD$^{2}$. The vertical scales are in the same arbitrary units in all graphs. Note that the pics at 1290 nm and 1540 nm could arise from other quantum dots, since they display similar saturation behavior with increasing power.}
        \label{Fig:SpectrumSingleDots}
\end{figure}

In order to investigate in more detail the characteristics of the islands corresponding to different parts of the spectrum, we examined two emission lines, corresponding to the excitons of two different quantum dots, referred to as QD$^{1}$ (1298 nm) and QD$^{2}$ (1538 nm). 
As QD$^{2}$ lies in a relatively dense part of the spectrum, it was studied on a smaller mesa of diameter of 0.5 $\mu$m, containing less than 20 islands far enough from the edge so as to luminesce. 
A filter cutting-off wavelengths below 1500 nm was used.
The photoluminescence spectra of the two mesas at 4 K at different incident powers
under pulsed excitation, are shown on Figure \ref{Fig:SpectrumSingleDots}. 
At low incident powers we observe sharp peaks that may be attributed to the emission from the exciton and the s-shells of the QD$^{1}$ and QD$^{2}$ quantum dots. 
These peaks lie on a broad background, whose intensity increases with increasing incident power.
The intensity of the background is approximately twice as high around QD$^2$ as compared with QD$^1$, because of the higher density of quantum dots at that wavelength.
In order to confirm that the background signal arises from the islands themselves and not from the surrounding materials (wetting layer, impurities, defects...), a similar sample with just the wetting layer and no islands was grown and subsequently processed into mesas under the same conditions. No luminescence was observed on this sample at wavelengths longer than the spectral band edge of the wetting layer (around 1.15 $\mu$m i.e. 1.08 eV) at any incident power. 

When the incident power is increased, we observe the emergence of additional lines attributable to p-shell emission. For QD$^1$ they are around 1250 nm to 1260 nm, giving a spacing between the s- and p-shells of $\sim$30 meV. 
The p-shell of QD$^2$ corresponds to a group of peaks around 1505 nm, giving a spacing between the s- and p-shells of at least $\sim$18 meV for QD$^{2}$. It should be noted that at low incident intensities additional lines are present around 1510 nm, due to excitons from other quantum dots.
Assuming that the islands correspond to rectangular potential wells for electron-hole pairs and that the p-shells involve only the two quantum numbers in the plane perpendicular to the growth direction, these values indicate that the bottom of the rectangular well (i.e.\ the effective bandgap of a layer confined in the growth direction to the height of the QDs) is at $\sim$935 meV for QD$^{1}$ and at $\sim$800 meV for QD$^{2}$. 
Since the bandgap of InAs$_{0.8}$P$_{0.2}$ is at 600 meV, if we assume that difference between QD$^{1}$ and QD$^{2}$ is due to confinement in the growth direction, the values of the effective bandgaps would indicate that QD$^{2}$ is 30 \% taller than QD$^{1}$ .

While we conclude that the s-p shell spacing is of the order to 30 meV for QD1, this assertion would be strongly supported by a cross-correlation measurements. The latter was not possible due to the low intensity of these high-energy emission lines, requiring extremely long integration time. However the power dependence of their intensity give some insight on their nature. As expected for p-shell states, these emission lines only appear above the saturation of the exciton line and all associated sharp lines of the p-shell grow in a similar way.  Their integrated emission does not saturate and follows closely the mean number of electron-hole pairs as function of the pump power from figure 6. Also the energy spacing is of the rigth order of magnitude when compared to theoretical predictions \cite{Gong08}

The position of the wetting layer bandgap at approximately 1.08 eV indicates that QD$^{1}$ corresponds to a well-depth of about 150 meV accommodating some 8 electron-hole levels (assuming a two-dimensional rectangular-well potential), while QD$^{2}$ has a well-depth of about 280 meV, with possibly at most 35 electron-hole levels. This approach is valid for quantum dots with relatively large lateral extension which is the case in the system under study.
Recent STM measurements in cleaved InAsP/InP quantum dots \cite{Fain2010,Fain2011} permitted the mapping of electron wavefunctions with as many as 5 nodes along the base of the dot in the exposed surface. Assuming a similar range of values for the quantum number in the perpendicular direction in  an uncleaved dot, that would correspond to more than 25 electron levels, thus corroborating the possibility of having a few tens of electron-hole levels in the quantum dot.
At low incident powers, only the lowest electron-hole level (exciton) is populated, giving rise to a single sharp line upon recombination. At higher incident powers, more than one electron-hole pair is injected in the quantum dot; emission from these multi-exciton states gives rise to the broad background \cite{Winger09}. The valence band offset for InAsP/InP quantum dots is higher than InAs/GaAs quantum dots \cite{Vurgaftmana2001}, almost equally distributed between electron and holes. In conjunction with the STM images \cite{Fain2010,Fain2011} a multi-level description of the quantum dot is consequently possible.

\begin{figure}[!ht]
    \centering
    \begin{tabular}{rr}
        \includegraphics[width=8cm]{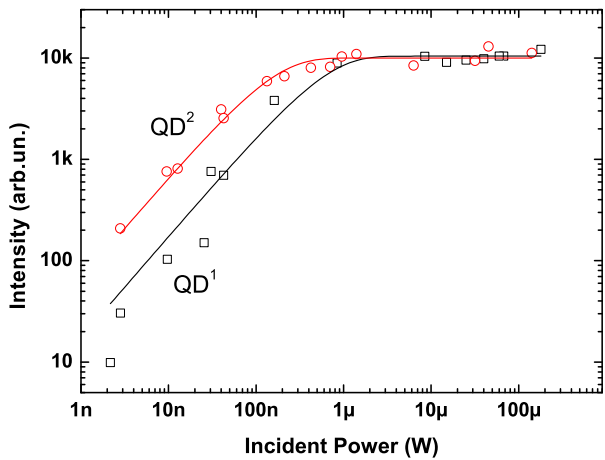} \\
        \end{tabular}    
    \caption{Intensity of the exciton peak of QD$^{1}$ (open black squares) and QD$^{2}$  (open red circles) as a function of incident power. Curves are fits to Eq.\ (\ref{eq:saturation}) with P$_{sat}$ = 620 nW for QD$^{1}$ and P$_{sat}$ =180 nW for QD$^{2}$.}
        \label{Fig:IntegratedSpectrumSingleDots}
\end{figure}

The dependence of the intensity of the exciton emission lines for QD$^{1}$ and QD$^{2}$ on the incident power is presented on Figure \ref{Fig:IntegratedSpectrumSingleDots}, and is reasonably well described by a saturation formula of the form 
\begin{equation}
I_{out} = I_0 (1-e^{-P_{in}/P_{sat}})
\label{eq:saturation}
\end{equation}
which assumes Poisson statistics for the number of electron-hole pairs in the quantum dot, with $P_{sat}$, the incident power at saturation, corresponding to the input power for which one electron-hole pair is injected on average in the quantum dot. 
Its value is 620 nW for QD$^{1}$ and 180 nW for QD$^{2}$.
At low incident powers, the intensity of these lines increases approximately linearly as it is proportional to the probability of trapping a single electron-hole pair in the quantum dot. 
Beyond $P_{sat}$ the intensity saturates to a constant level, as more than one electron-hole pair is trapped in the quantum dot during each excitation cycle: the multi- electron-hole pair state contributes to the broad background \cite{Winger09}, as the pairs recombine successively, and only the last pair in the cascade contributes to the sharp emission line. Note that it is not simple to differentiate a neutral exciton from a charged exciton, and precise identification requires polarization resolved quasiresonant excitation\cite{Laurent2005}. Eventually, let's note the unexpected absence of biexciton lines, which is currently under investigation.

%On the other hand, the background continuous to grow but increases sublinearly (figure \ref{Fig:IntegratedSpectrumSingleDots}. Assuming that the background is due to the excess of electron-hole pairs that are captured by the high excitated states of the QD, the background is expected to grow sulinearly, as we will explain later in the text.

%When increasing temperature, QD$^{1}$ and QD$^{2}$ are red-shifted and their
%linewidths, measured at an excitation power of 1.45$P_{sat}$ and X$P_{sat}$
%respectively, display a linear dependence on temperature up to 60 K:
% \begin{equation}
%\Gamma(T)=\Gamma_0+\alpha T 
%\end{equation}
%where $\Gamma_0$ is the linewidth at 0 K. The linear term $\alpha T$ can be
%attributed to linewidth broadening due to acoustic phonon scattering. 
%For both lines investigated, we obtained $\Gamma_0=39 \pm 5$ $\mu$eV and $\alpha=6.8
%\pm 0.5$ $\mu$eV.K$^{-1}$, in the same range as the values reported for InAs/GaAs
%quantum dots \cite{Favero2007}. However, in contrast to InAs/GaAs quantum dots, we
%do not observe any additional broadening due to optical phonon scattering, up to a
%temperature of 60K.

\section{Time resolved measurements}
\label{TRM}

%Another signature of the ``quantum dot'' nature of these emitters is multi-exciton cascade emission. Such cascade can be observed by measuring the emission line decays as a function of excitation power. 

\begin{figure}[!ht]
     \centering
       \begin{tabular}{c}
        \includegraphics[width=7.5cm]{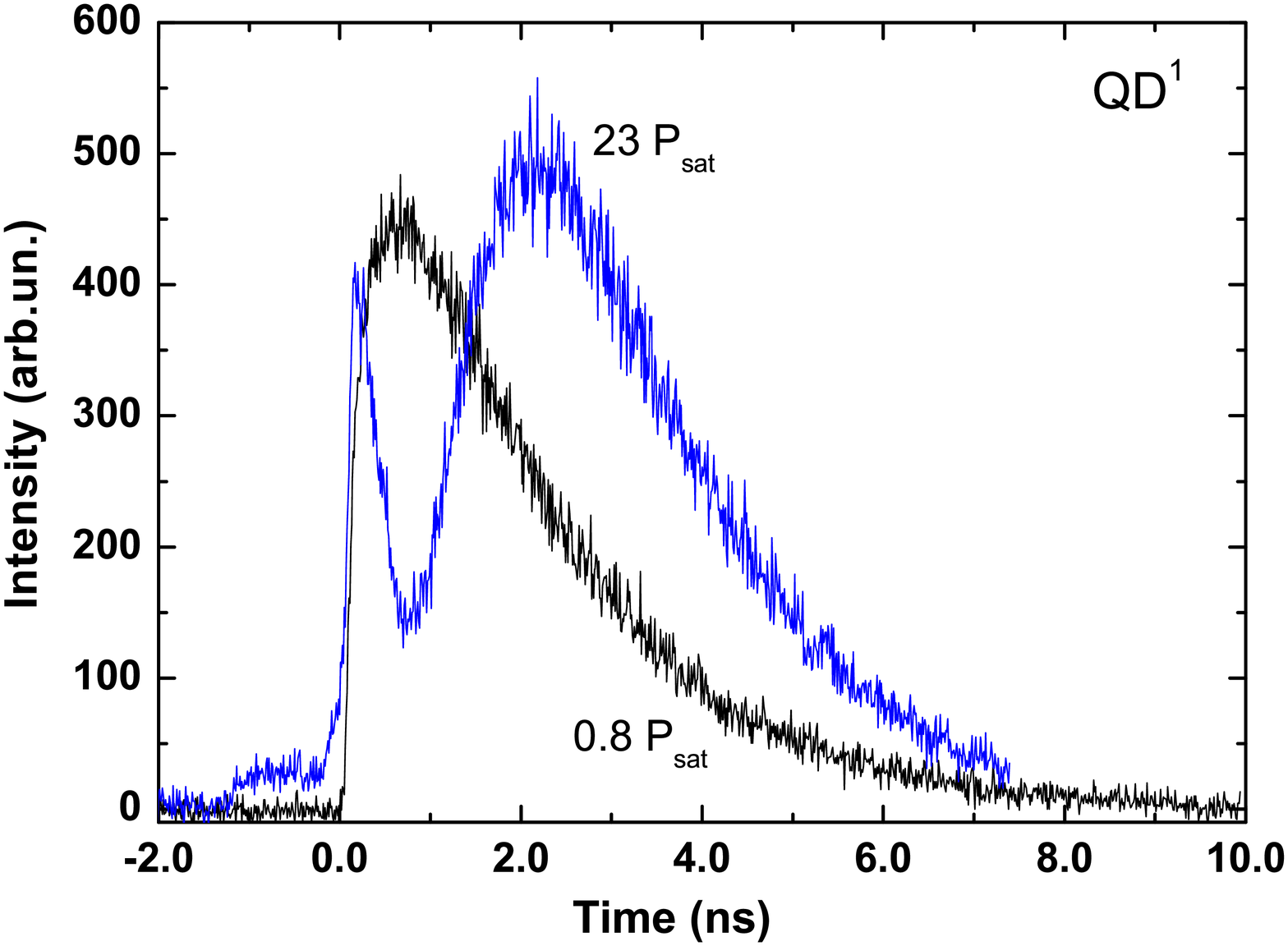} \\
        \includegraphics[width=7.5cm]{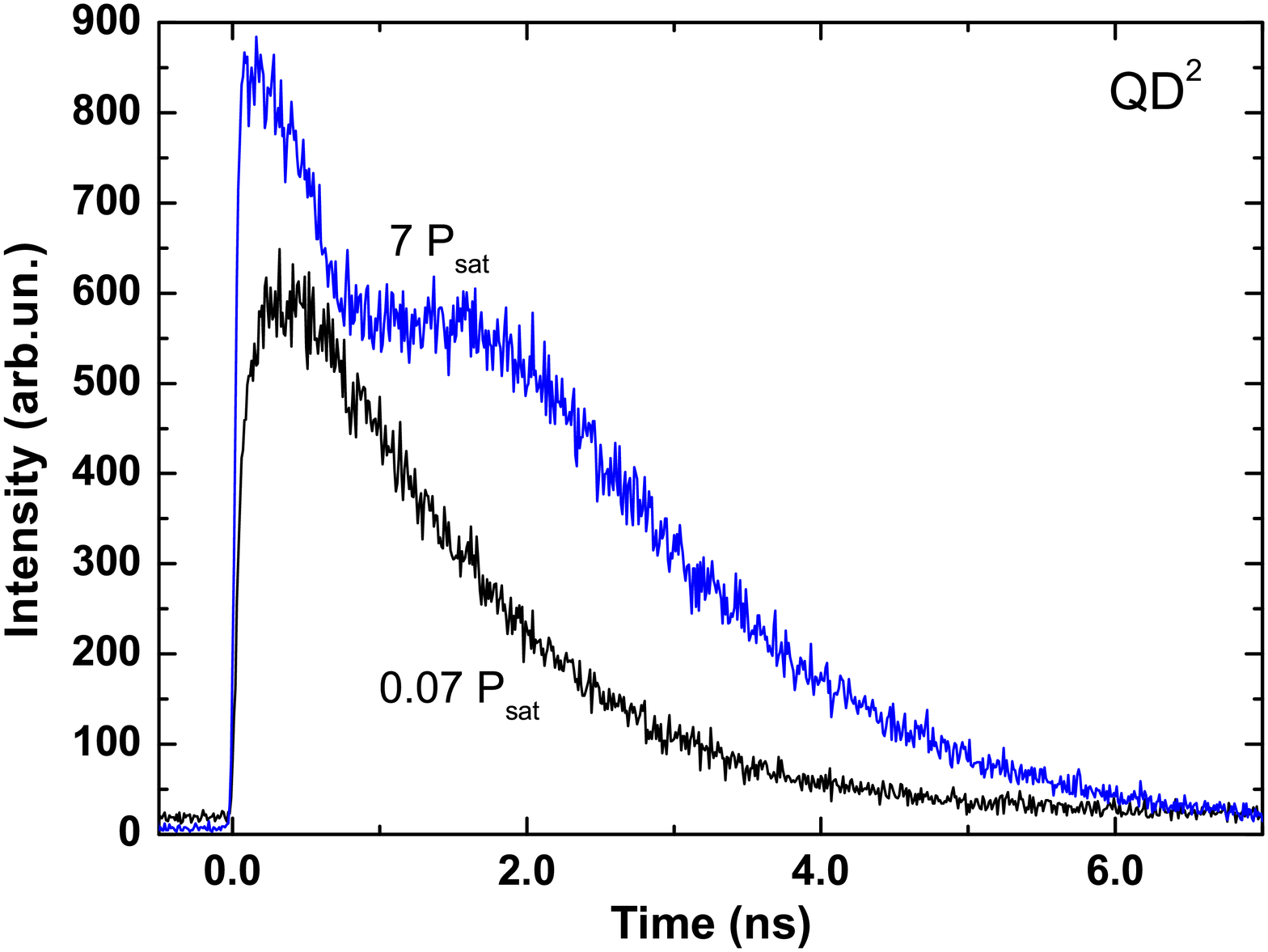} 
        \end{tabular} 
    \caption{Decay curves of the exciton emission lines associated to two
different InAsP/InP quantum dots labeled QD$^{1}$ (Top) and QD$^{2}$ (Bottom) obtained under pulsed excitation at 840 nm for different incident powers. (Top) QD$^1$: P$_{in}$ = 0.8 P$_{sat}$ (black curve) and P$_{in}$ = 23 P$_{sat}$ (blue curve) (Bottom) QD$^2$: P$_{in}$ = 0.07 P$_{sat}$ (black curve) and P$_{in}$ = 7 P$_{sat}$ (blue curve). Note P$_{sat}$ = 620 nW for QD$^{1}$ and P$_{sat}$ =180 nW for QD$^{2}$}
        \label{Fig:DecaySingleDots}
\end{figure}

The formation and recombination dynamics of the quantum dot excitons was investigated by monitoring the time dependence of the exciton emission line under pulsed excitation.
When the laser pulse is absorbed, it generates a large number of electron-hole pairs in the InP buffer and the wetting layer, some of which are captured by the quantum dot, giving rise to its luminescence upon recombination. 
As the unexcited quantum dots may already contain one or two electrons due to the residual doping, the capture of carriers may proceed charge-by-charge, in the weak excitation regime \cite{Kapon2006}. 
The electrons generated by the absorption of the incident pulse remain in the buffer, as they are attracted to the ionized donors, while the holes are efficiently captured by the negatively-charged quantum dots, where they eventually recombine radiatively with the resident electrons to produce the quantum dot emission. 
Thus, at low incident powers, the time dependence of the QD$^{1}$ and QD$^{2}$ exciton lines display a sharp rise limited by the temporal resolution of our setup of 70 ps, indicating that the exciton state is populated immediately through the capture of a hole (when the quantum dot already contains one electron) or the capture of an electron-hole pair (when the quantum dot is empty).
This immediate response is followed by a rapid buildup of the exciton intensity (with an exponential characteristic time of about 400 ps), up to a maximum at $\Delta t =$ 500 ps later for QD$^{1}$ and $\Delta t =$ 200 ps for QD$^{2}$ (see Fig.\ \ref{Fig:DecaySingleDots}). 
This delay indicates that the charge capture populates also a relay state which in turn feeds the exciton.
As the relative weight of the relay state is of the same order as that of the exciton, even for incident intensities as low as 2\% of saturation, it is quite likely that the relay state corresponds to the biexciton produced by the capture of
two holes when the quantum dot already contains two electrons. 
Thus, at low incident intensities, the relative weights of the prompt and delayed components reflect essentially the doping statistics \cite{Brescol2011}.
After reaching the maximum, the luminescence decays exponentially with a characteristic time of 2.2 ns for QD$^{1}$ and 1.4 ns for QD$^{2}$ (see Fig. \ref{Fig:DecaySingleDots}), which correspond to the exciton lifetimes $\tau_X$ in the two quantum dots.

%It should be noted that, under standard growth conditions, the residual doping of the InP growth is measured around $10^{16} e^- cm^{-3}$, each quantum dot may contain a few electrons, when unexcited, which would be matched by the
%capture of an equal number of holes upon excitation \cite{Kapon2006} to produce the
%emission of the neutral exciton monitored here. This would skew the Poisson
%statistics assumed in Eq.\ref{eq:saturation}. Our data would be compatible with such a
%charge-by-charge capture process, but their precision does not permit us to
%discriminate between these two options.

%We note that the decay time is constant for QD$^{1}$ up to a temperature of 80K, indicating a very small contribution of thermally activated decay paths.

%For QD$^{3}$, the decay curve is a bi-exponential resulting from two contributions:
%the QD$^{3}$ emission line itself and the broad background.
%In order to obtain the decay curve of QD$^{3}$ itself, we measured the background
%decay in the vicinity of QD$^{3}$ (corresponding to an exponential decay of 900 ps)
%and then subtracted its contribution from the overall decay curve. 
%The QD$^{3}$ decay was thus found to be an exponential of 1.3 ns. It was not,
%however, possible to ascertain whether there was a relay level in this case also.

When the incident power is increased, the decay curves change in two ways: First, the delay $\Delta t$ between the incident pulse and the maximum of the exciton intensity increases and, second, an additional fast decay emerges at short time delays, with a characteristic time of the order of 400 ps and 880 ps for QD$^{1}$ and QD$^{2}$ respectively. 
The increasing delay $\Delta t$ results from the injection of increasingly more electron-hole pairs in the island, which must decay before emission can occur from the one-exciton level that is being monitored.
The fast component, on the other hand, is due to the emergence of the broad background whose contribution increases with incident power (see Fig. \ref{Fig:SpectrumSingleDots}). 
This is confirmed by measuring the temporal traces of the background signal in the vicinity of QD$^{1}$ and QD$^{2}$: the decay times are respectively 400 ps and 800 ps.

\begin{figure}[!ht]
     \centering
        \begin{tabular}{c}
         \includegraphics[width=7.5cm]{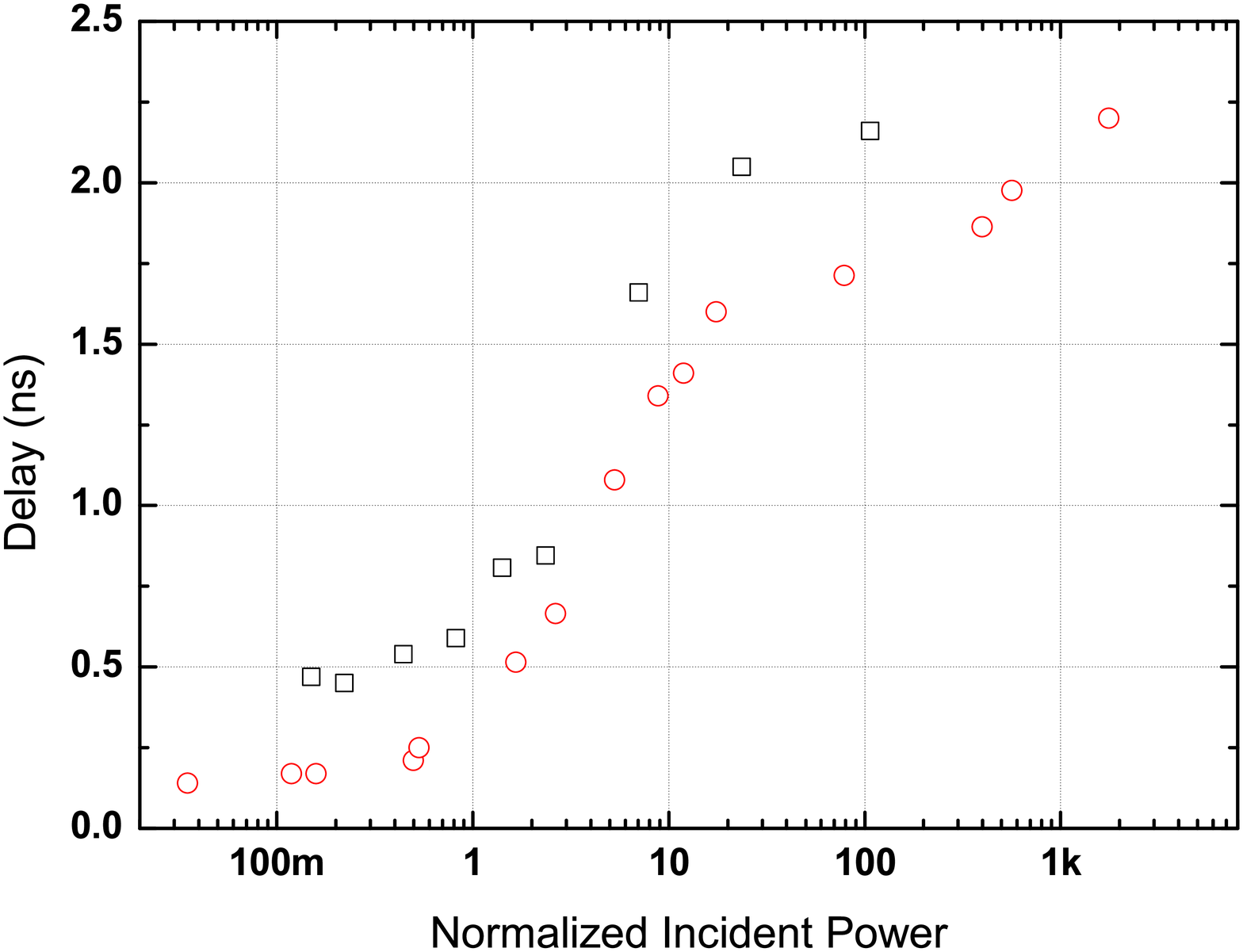}\\
        \includegraphics[width=7.5cm]{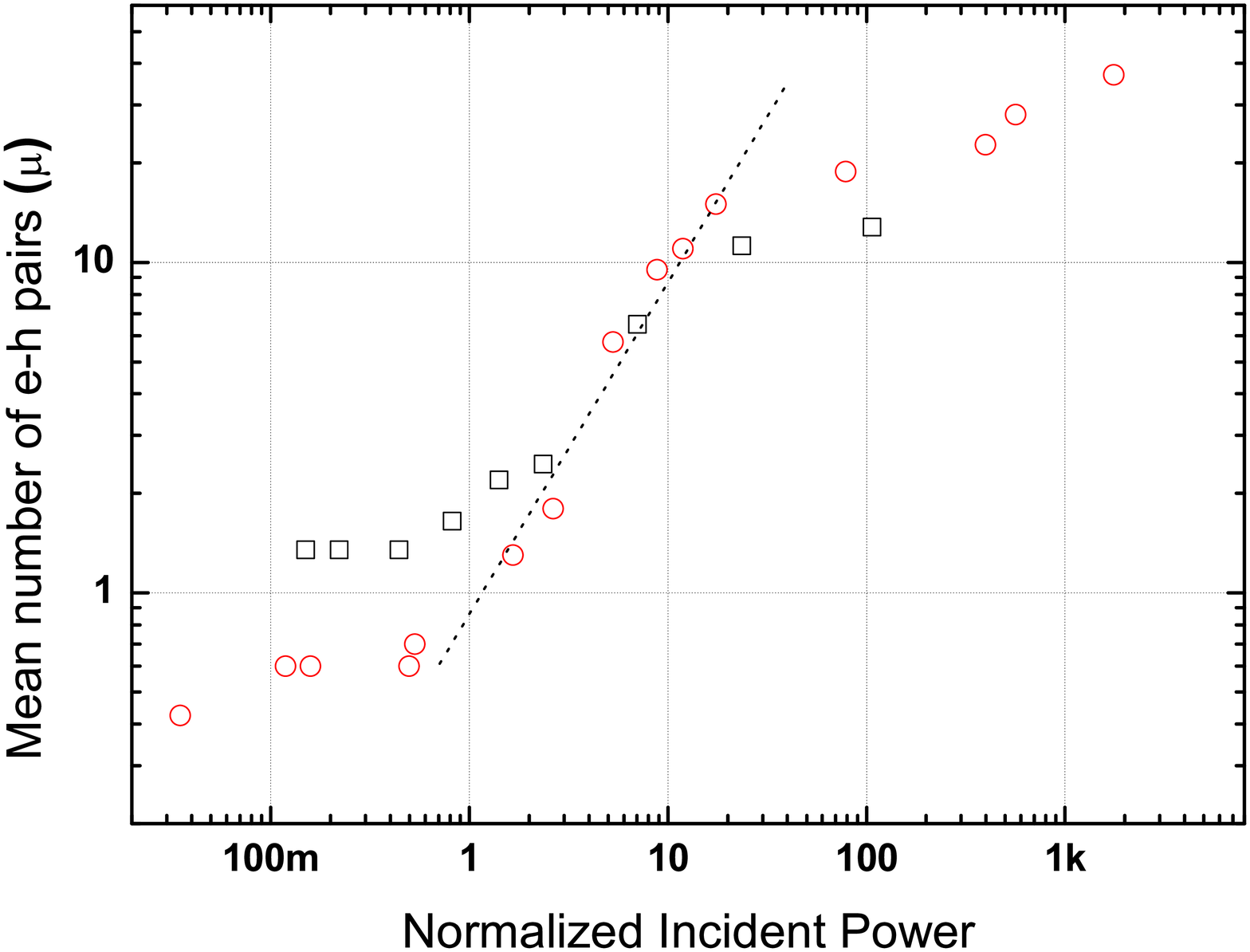}
        \end{tabular}  
    \caption{
(Top) Time delay between the incident excitation pulse and the intensity maximum of the exciton luminescence peak of QD$^{1}$ (open black squares) and QD$^{2}$  (open red circles) as a function of normalized incident power under pulsed excitation at 840 nm, in a log-linear scale. The normalization is with respect to P$_{sat}$ = 620 nW for QD$^{1}$ and P$_{sat}$ = 180 nW for QD$^{2}$. (Bottom) Mean number of electron-hole pairs ($\mu$) in QD$^{1}$ (open black squares) and QD$^{2}$ (open red circles) as a function of incident power under pulsed excitation at 840 nm, in a log-log scale. 
The mean number of electron-hole pairs is obtained by fitting the exciton decay curves using Eq.\ (\ref{eq:delay2}) for QD$^{1}$ and Eq.\ (\ref{eq:delay1}) for QD$^{2}$. The dotted line is a guide for the eye with unity slope. }
        \label{Fig:muSingleDots}
\end{figure}

Figure \ref{Fig:muSingleDots} (Top) shows the delay of the exciton line $\Delta t$ for both QD$^1$ and QD$^2$ as function of the normalized incident power. 
As this delay is directly related to the mean number of electron-hole pairs in the dot ($\mu$), we can deduce $\mu$ from the time-resolved curves (see Appendix A). 
A plot of $\mu$ as a function of the normalized incident power is given in Fig.\ \ref{Fig:muSingleDots} (Bottom). 
These values were obtained by fitting the time-resolved curves using Eq.\ (\ref{eq:delay2}) for QD$^1$ with $\tau_X=2.2$ ns and $\tau_B=0.72$ ns and using Eq.\ (\ref{eq:delay1}) for QD$^2$ with $\tau_X=1.4$ ns and $\tau_B$=0.9 ns. The exciton lifetime is directly measured from the curves, while $\tau_B$ and $\mu$ are fitting parameters.
The slight discrepancy between these values for $\tau_B$ and those obtained directly from the decay of the background, is due to the fact that the former involve only those cascades that end up in the one-exciton line that is being monitored, while the latter involve all transitions at the corresponding wavelength.

At low incident powers, the mean number of electron-hole pairs present in the quantum dot is constant, corresponding to $\mu\approx$1.1 for QD$^1$ and  $\mu\approx$0.5 for QD$^2$, reflecting essentially the mean number of electrons resident in each dot due to the residual doping. This gives rise to the constant delay observed in the time resolved curves at low excitation power.
Under these conditions, emission results from injecting in the dot a matching number of holes, charge-by-charge.
At higher incident intensities, the number of captured charges exceeds the number of resident electrons and the average number of electron-hole pairs increases with the incident power, initially linearly and then sublinearly. 
Deviation from linearity and saturation of $\mu$ are visible beyond 10 P$_{sat}$.
The reason is that under high incident power, a dense electron-hole gas is generated in the InP buffer layer so that bimolecular radiative recombination enters in competition with pair capture by the quantum dot: an increase of the electron-hole population in the InP buffer gives rise to a quadratic increase radiative recombination rate inside the buffer layer and therefore a sub-linear increase in the pair capture by the quantum dot, as discussed in Appendix B.

\begin{figure}[!ht]
     \centering
        \begin{tabular}{c}
         \includegraphics[width=7.5cm]{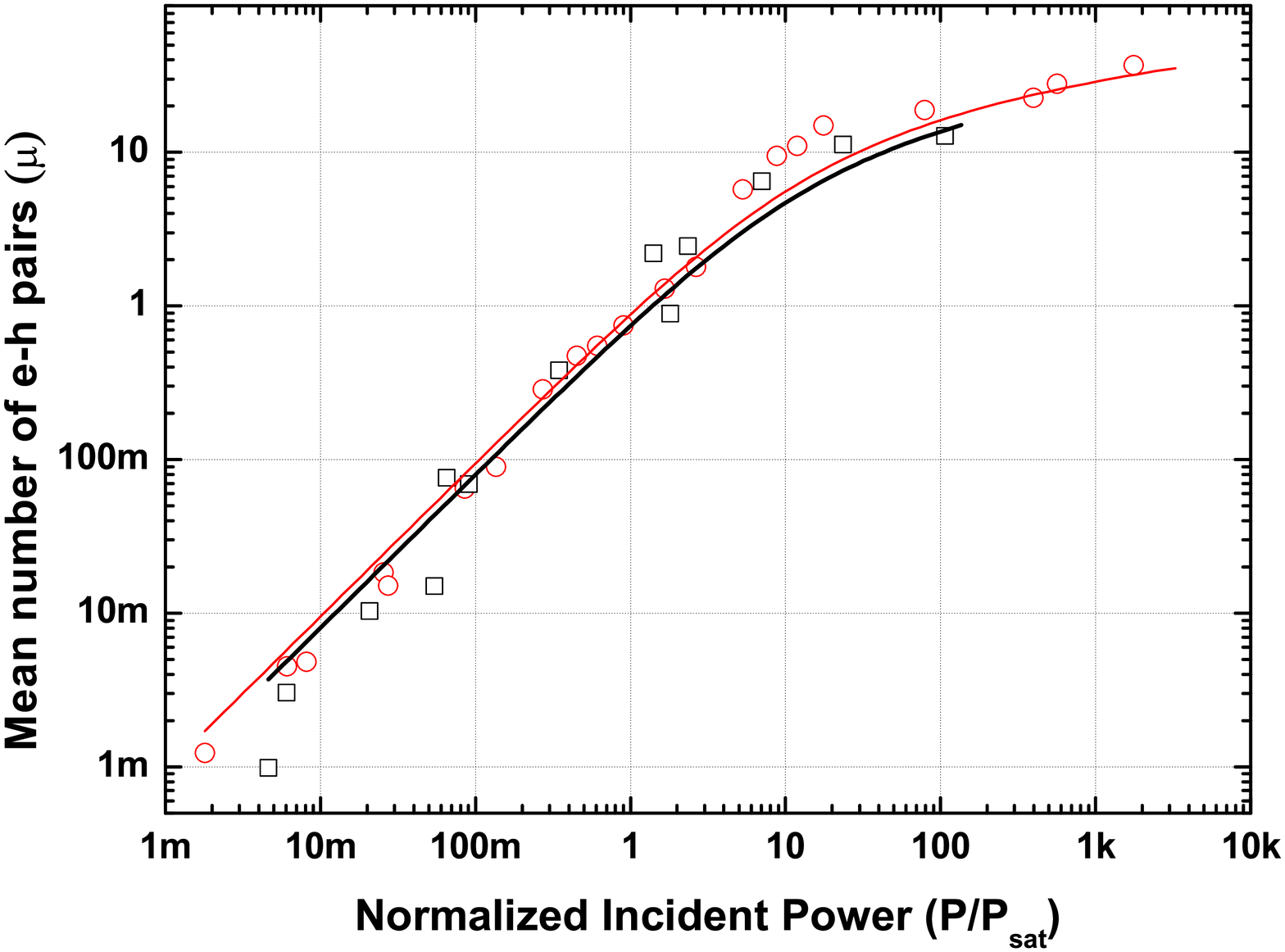} \\
         \includegraphics[width=7.5cm]{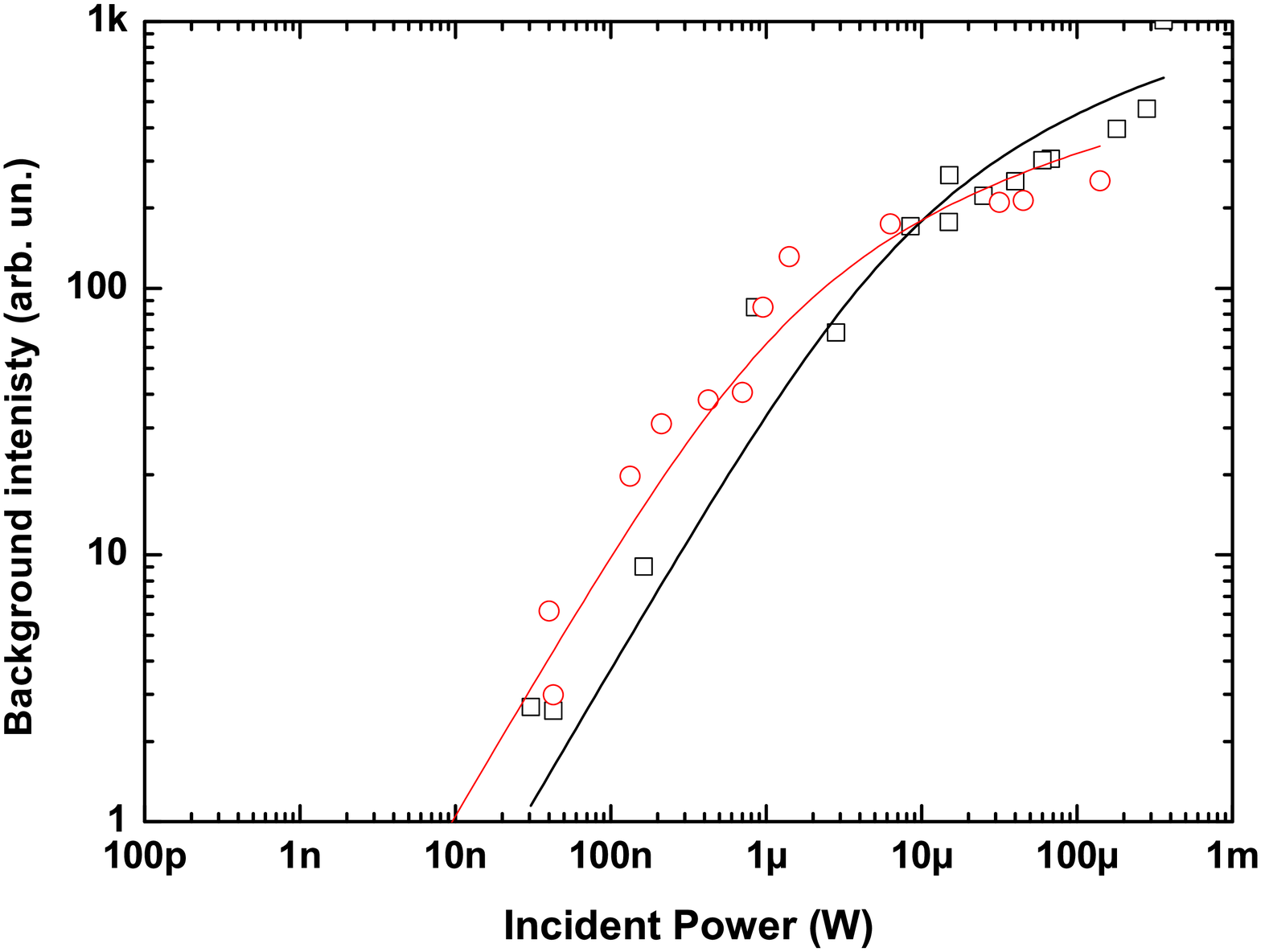}
        \end{tabular}  
    \caption{(Top) Mean number of electron-hole pairs ($\mu$) injected in QD$^{1}$ (open black squares) and QD$^{2}$ (open red circles) as function of the normalized excitation power. 
        (Bottom) Intensity of the background signal in the spectral vincinity of QD$^{1}$ (open black squares) and QD$^{2}$  (open red circles) as function of incident power under pulsed excitation at 840 nm, in log-log scale. All fits are obtained using Eq.\ (\ref{eq:logcapture2}).}
        \label{Fig:DelaySingleDots}
\end{figure}

As the statistics of injected pairs is masked by the doping statistics at low incident powers, we can compile a composite set of data for the mean number of injected pairs by using the intensity data of Fig.\ \ref{Fig:IntegratedSpectrumSingleDots} at low incident powers (below $P_{sat}$) and the data of Fig.\ \ref{Fig:muSingleDots} (Bottom) at high incident powers (above $P_{sat}$). 
The resulting set of data is shown on Fig.\ \ref{Fig:DelaySingleDots} (Top), and is quite well fitted by Eq.\ (\ref{eq:logcapture2}), using VC/B = 4.7 for QD$^1$ and VC/B = 5.6 for QD$^2$. 
This analysis indicates that only a small number of electron-hole pairs reaches the dot, even at very high incident intensities, because the excitation process involves the generation of an electron-hole gas in the buffer layer and a subsequent capture of electron-hole pairs by the quantum dot: for incident intensities of the order of $P_{in}=1000 P_{sat}$ the quantum dot contains only 20 to 25 pairs. Such a high number of levels is indeed expected from the STM experiments on equivalent quantum dots \cite{Fain2010,Fain2011}.
During the radiative cascade of a multiply-excited quantum dot, only the last electron-hole pair emits in the exciton line, while all others contribute to the broad background. This means that the background intensity should be proportional to the number of pairs present in the dot.
Fig.\ \ref{Fig:DelaySingleDots} (Bottom) shows the background intensity as function of incident power in the vicinity of QD$^1$ and QD$^2$. 
The curves are a fit using Eq.\ \ref{eq:logcapture2} with the same parameters as for the mean number of pairs above, plus an overall vertical scaling parameter to account for the arbitrary units of the intensity.
The fit is quite good, considering that the background contains also contributions from other quantum dots, each with a different saturation power.
Thus, this fit supports the assumption that the broad background emission arises directly from the electron-hole pairs captured by the quantum dot \cite{Winger09}.

\section{Photon correlation measurements}

\begin{figure}[!ht]
    \centering
    \includegraphics[width=8cm]{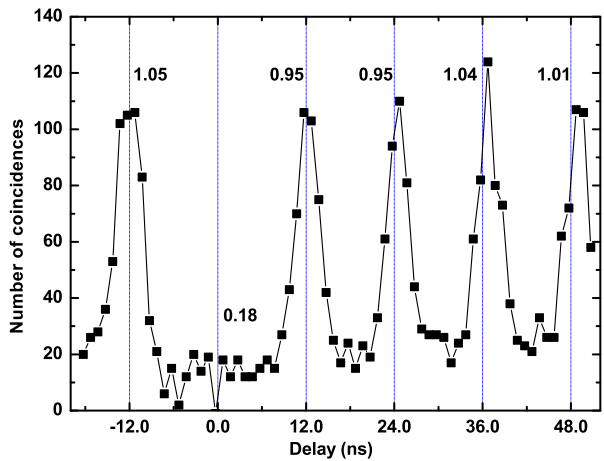}
    \caption{Histograms of the time intervals between detection events on the start and stop detectors for the QD$^{1}$ line under pulsed excitation with an incident power of 1.45 P$_{sat}$.}
        \label{Fig:Antibunching}
\end{figure}

Although cascade emission is a unique signature of confined 0D electronic states of single quantum dots, a direct proof of the unicity of the emitter is given by the observation of single-photon emission through anti-bunching. 
Photon correlation measurements were performed by measuring the normalized second-order correlation function $g^{(2)}(\tau)=<:I(t)I(t+\tau):>/<I(t)>^2$ where $I(t)$ is the emission intensity at time $t$ and $<::>$ indicates the normal ordering of the creation and annihilation operators.
Figure \ref{Fig:Antibunching} shows a histogram of the delay times between detection
events on the ''start'' and ''stop'' channels of the Hanbury-Brown and Twiss setup,
obtained for QD$^{1}$, at an incident power of 1.45 $P_{sat}$. 

Under pulsed excitation, the histogram of time delays consists of a series of peaks separated by the period of the pump cycle. 
When normalized by dividing by the count rates of the two detectors, by the
experiment time, and by the period of the pump cycle, and after subtracting the accidental coincidences due to the dark
counts, the area under each peak, except for the peak at $\tau=0$, is equal to 1 \cite{Beveratos02}.
The value of the histogram in the region around $\tau= 0$ measures the conditional
probability of detecting a second photon during the excitation cycle given that a
first photon has already been detected. It is expected to be below 0.5 for a single photon source. 
For QD$^{1}$, the normalized intensity level of the central region is 0.18.
This value is essentially equal to the ratio between the sharp line and the contribution of the broad background at that wavelength and at an incident power of $P=1.45 P_{sat}$, indicating that the QD$^{1}$ exciton is indeed a single photon emitter, and that the possibility of having a second photon is only due to the background emission. 
For QD$^{2}$, it was not possible to obtain the second order autocorrelation function because of the lower count rate and the strong contribution of the broad background.
 
%%%%%%%%%%%%%%%%%%%%%%%%%%%%%%%%%%%%%%%%%%%%%%%%%%%%%%%%%%%%%%%%%%%
\section{Summary and Conclusion}

In this paper, we have examined the photoluminescence characterisics of InAsP/InP quantum dots emitting at telecommunications wavelengths. 
Time-resolved measurements have shown that these dots can be multiply excited and undergo a radiative cascade. 
All but the last step of the radiative cascade give rise to a broad spectrum, while the last step, which involves emission from the one-exciton state, gives rise to a sharp spectral line.
The sharp line emission shows strong anti-bunching, indicating that InAsP/InP quantum dots are good single photon emitters, useful for quantum communications at telecommunications wavelengths. 
However, these quantum dots may not be suitable for some quantum information protocols, since the emission of the exciton and the biexciton lines will present a strong jitter due to the cascaded emission.
The broad background resulting from the radiative cascade is important for the use of these quantum dots as gain material in photonic crystal nanolasers, as the broad spectrum permits efficient feeding of a nanocavity that would otherwise be out of resonance with the quantum dot exciton. However, as the number of electron-hole pairs in the dot grows sub-linearly as a function of pumping, it is possible that the lasing threshold may appear to be progressive rather than abrupt.
\\

%%%%%%%%%%%%%%%%%%%%%%%%%%%%%%%%%%%%%%%%%%%%%%%%%%%%%%%%%%%%%%%%%%%
\noindent {\bf Acknowledgements:} 
The authors are thankful to Dr.\ T.\ Debuisschert for lending critical experimental equipment. 
This work was partly supported by the C'Nano Ile-de-France project ``CRYPHO" and the European project IST-249012 ``Copernicus''.
\\
%%%%%%%%%%%%%%%%%%%%%%%%%%%%%%%%%%%%%%%%%%%%%%%%%%%%%%%%%%%%%%%%%%%

\appendix
\section{Appendix A: Radiative Cascades}
\label{Appendix:Label}
In this Appendix, we examine the cascaded radiative decay of a multiply excited quantum dot.
We consider a quantum dot initially containing $N$ electron-hole pairs that recombine independently.
During the successive recombination events, the quantum dot will emit a cascade of $N-1$ photons into the broad
background until it reaches the one-exciton state whose emission is being monitored \cite{Santori2002, Laucht2010}. 
Assuming that the characteristic time for emission into the broad background $\tau_B$ is independent of the number of excitons present, the probability of reaching the two-exciton state at time $t$ is
\begin{equation}
P_N^2(t)= \frac{N (N-1)}{2} e^{- 2 t/\tau_B } \left(1-e^{- t/\tau_B} \right)^{N-2}
\label{eq:Probability2}
\end{equation}
Emission from this level feeds directly the exciton, whose luminescence corresponds to a sharp line and has a lifetime $\tau_X$.
The time-evolution of the exciton emission is thus given by the convolution of the feeding mechanism of Eq.\ (\theequation) with the exciton decay. 
The resulting summation of exponentials cannot be performed analytically for an arbitrary ratio $\tau_B/\tau_X$. 
Analytic approximations can nevertheless be obtained in two significant cases:

\noindent
(a) When $\tau_B/\tau_X \approx 1$, we have:
\begin{equation}
P_N^1(t)
\approx N e^{- t/\tau_X } \left(1-e^{- t/\tau_B} \right)^{N-1}
\label{eq:Probability1-1}
\end{equation}
and (b) when $\tau_B/\tau_X \ll 1$, we have:
\begin{widetext}
\begin{eqnarray}
P_N^1(t)
\approx e^{- t/\tau_X } \left( \left(1-e^{- t/\tau_B} \right)^N+ N e^{- t/\tau_B }
\left(1-e^{- t/\tau_B} \right)^{N-1} \right)
\label{eq:Probability1-0}
\end{eqnarray}
\end{widetext}

The number of excitons injected in a quantum dot follows, in principle, Poisson statistics.
However, a fraction of these excitons is non-radiant (dark) and do not participate in the cascade that leads to the radiant (bright) exciton level being monitored. 
The resulting statistics of the fully radiant multiexciton states involves hyperbolic Bessel functions rather than exponentials.
However, the limited precision of our experimental data does not permit us to distinguish between these statistics 
and those of the Poisson distribution which includes both radiant and non-radiant states.
Thus, for the sake of simplicity, we consider here Poisson statistics.

For a Poisson distribution of the number of electron-hole pairs $N$ with mean $\mu$, and for the case $\tau_B/\tau_X \approx 1$, convolution of Eq.\ (\ref{eq:Probability1-1}) with the exciton decay gives the exciton intensity as a function of time as:
\begin{equation}
I_X(t)=  \mu e^{-t/\tau_X} e^{-\mu \exp[-t/\tau_B]}
\label{eq:delay1}
\end{equation}
with a maximum at 
\begin{equation}
\Delta t= \tau_B \ln \left ( \mu \frac{\tau_X}{\tau_B} \right )
\label{eq:deltat1}
\end{equation}
while for the case $\tau_B/\tau_X \ll 1$, the convolution of Eq.\ (\ref{eq:Probability1-0}) gives the exciton intensity as:
\begin{equation}
I_X(t)= e^{-t/\tau_X} \left ( e^{-\mu \exp[-t/\tau_B]} (1+ \mu e^{- t/\tau_B})-e^{-\mu } \right )
%I_X(t)= e^{-t/\tau_X} \left ( e^{-\mu e^{-t/\tau_B}} (1+ \mu e^{- t/\tau_B})-e^{-\mu } \right )
\label{eq:delay2}
\end{equation}
with a maximum at 
\begin{equation}
\Delta t \approx \tau_B \ln \left( \mu \sqrt{\frac{\tau_X}{\tau_B}} \right).
\label{eq:deltat2}
\end{equation}

\section{Appendix B: Carrier Capture}

We consider the process whereby an incident light pulse is absorbed in a thick semiconductor layer, thus generating a density $n$ of electron-hole pairs, which may either recombine radiatively or be captured by the quantum dots embedded in the semiconductor layer.
The kinetics of the carriers in the semiconductor layer are described by:
\begin{equation}
\frac{dn}{dt} = -A n - B n^2 ,
\label{eq:kinetics}
\end{equation}
where $A$, the monomolecular decay rate, includes capture of the electron-hole pairs by the quantum dots, radiative recombination involving the residual doping, as well as non-radiative recombination of the carriers. 
$B$ is the familiar bimolecular radiative recombination coefficient.
This differential equation has the well-known solution
\begin{equation}
n(t) = \frac{A n_0 e^{-A t}}{A+ B n_0 (1-e^{-A t})} ,
\label{eq:soln}
\end{equation}
where $n_0$ is the initial pair density produced by the absorption of the incident light pulse at $t=0$.
As carrier capture by the quantum dot is much faster than the radiative lifetime of the pairs in the dot (typically by 3 orders of magnitude) the kinetics of pair capture in the dot may be written as 
\begin{equation}
\frac{d \mu}{dt} = V C n ,
\label{eq:capture}
\end{equation}
where $\mu$ is the mean number of pairs in the dot, C is the capture rate per quantum dot and V is the volume of the quantum dot. 
Thus, the number of pairs captured by the dot may be written as:
\begin{equation}
\mu = V C \int_0^{\infty} n(t) dt = \frac{V C}{B} \ln \left( 1+ \frac{B}{A} n_0 \right). 
\label{eq:logcapture}
\end{equation}
In our experiments, we do not have direct access to $n_0$, but we can assume that it is proportional to the incident pulse energy and thus to the mean incident power, P$_{in}$.
The proportionality constant may be obtained by considering the number of captured pairs in the low-density limit of Eq.\ (\ref{eq:logcapture}), 
\begin{equation}
\mu_{n_0 \rightarrow 0} = \frac{V C}{A} n_0 
\label{eq:lowdens}
\end{equation}
which, in Eq.\ (\ref{eq:saturation}) was assumed to be given by P$_{in}$/ P$_{sat}$.
Thus, the initial pair density in the excited semiconductor layer can be written as
\begin{equation}
n_0 = \frac{A}{V C}  \frac{P_{in}}{P_{sat}}
\end{equation}
so that the mean number of electron-hole pairs in the quantum dot as a function of the incident power reads
\begin{equation}
\mu = \frac{V C}{B} \ln \left( 1+ \frac{B}{V C} \frac{P_{in}}{P_{sat}}  \right). 
\label{eq:logcapture2}
\end{equation}
Thus, while at low incident powers Eq.\ (\ref{eq:logcapture2}) reduces to the familiar proportionality between the number of pairs captured by the quantum dot and those injected in the wetting layer, at high incident powers the competition between the capture process and the radiative bimolecular recombination causes the number of pairs in the quantum dot to increase very slowly as a function of incident power.

Eq.\ (\theequation) can be used to fit our experimental data, with $\frac{VC}{B}$ as the only fitting parameter. 
An order of magnitude estimation of its value may be obtained by considering the orders of magnitude of the three parameters composing it:
B $\sim 10^{-10}$ cm$^3$.s$^{-1}$, V $\sim 10^{-18}$ cm$^3$.
The capture rate in multi-dot systems has been measured to be \cite{Fiore2000} A $\sim 10^{11}$ s$^{-1}$, which would correspond to a capture rate of C $\sim 10^8 - 10^9$ s$^{-1}$ per dot.
Thus $\frac{VC}{B} \sim 1 - 10 $.

\end{document}